\documentclass[]{emulateapj}
\slugcomment{}

\shorttitle{Nuclei and dwarfs}
\shortauthors{De Propris et al.}

\begin{document}


\title{A Comparison of surface brightness profiles for ultra-compact dwarfs 
       and dwarf elliptical nuclei: implications for the `threshing' scenario}


\author{R. De Propris and S. Phillipps}
\affil{H. H. Wills Physics Laboratory, University of Bristol,
       Tyndall Avenue, Bristol, BS8 1TL, UK; R.DePropris@bristol.ac.uk,
       S.Phillipps@bristol.ac.uk}

\author{M. J. Drinkwater}
\affil{Department of Physics, University of Queensland, Brisbane,
       QLD, 4072, Australia; mjd@physics.uq.edu.au}

\author{M. D. Gregg\altaffilmark{1}}
\affil{Department of Physics, University of California at Davis,
       and Institute for Geophysics and Planetary Physics, Lawrence
       Livermore National Laboratory; gregg@igpp.ucllnl.org}

\author{J. B. Jones}
\affil{School of Mathematical Sciences, Queen Mary University
       of London, Mile End Road, London, E1 4NS, UK; bryn.jones@qmul.ac.uk}

\author{E. Evstigneeva}
\affil{Department of Physics, University of Queensland, Brisbane,
       QLD, 4072, Australia; katya@physics.uq.edu.au}

\and

\author{K. Bekki}
\affil{Department of Astrophysics, University of New South Wales,
       Kensington, NSW, 2052, Australia; bekki@bat.phys.unsw.edu.au}


\altaffiltext{1}{also at Institute of Geophysics and Planetary Physics,
Lawrence Livermore National Laboratories, L-413, Livermore, CA 94550, USA}


\begin{abstract}

Using imaging from the Hubble Space Telescope, we derive surface
brightness profiles for ultra-compact dwarfs in the Fornax cluster and
for the nuclei of dwarf elliptical galaxies in the Virgo cluster.
Ultra-compact dwarfs are more extended and have higher surface
brightnesses than typical dwarf nuclei, while the luminosities,
colors, and sizes of the nuclei are closer to those of Galactic
globular clusters.  This calls into question the production of
ultra-compact dwarfs via ``threshing'' whereby the lower surface
brightness envelope of a dwarf elliptical is removed by tidal
processes, leaving behind a bare nucleus.  Threshing may still be a
viable model if the relatively bright Fornax ultra compact dwarfs
considered here are descended from dwarf ellipticals whose nuclei are
at the upper end of their luminosity and size distributions.

\end{abstract}




\section{Introduction}

It was noted almost 40 years ago that galaxies appear to define a 
relatively narrow region of the luminosity-surface brightness plane 
\citep{arp65}. Later on, \citet{disney76} postulated that this, rather 
than being the result of a real physical correlation, was due to a 
selection effect, where large diffuse objects, would be lost due to 
the brightness of the night sky, while small compact galaxies would 
be mistaken for stars and not included in redshift surveys. The existence 
of this correlation, or otherwise,has important implications for 
galaxy formation models \citep{dalcanton97,mo98}.

A small number of large, luminous low surface brightness galaxies have
been discovered \citep{bothun87,impey88}, but it is commonly held that
such systems are actually rare \citep{cross01}.  There
are also examples of very compact dwarf galaxies (e.g. POX 186 -- Kunth et al.
1988) but, since these objects are easily confused with stars in typical
imaging conditions, large `blind' redshift surveys are needed to ascertain 
their presence, and, as stars outnumber galaxies at the typical apparent 
magnitudes accessible to survey spectrographs, this is uneconomical in
terms of the telescope time needed.

\citet{drinkwater99}, however, showed that in the cluster environment the 
density of galaxies may be high enough, over a small field of view, that a 
redshift survey of all objects in the cluster area may be feasible, with 
acceptable rates of stellar contamination, in order to search for compact 
objects \citep{drinkwater00,phillipps01}. This effort was rewarded by the 
discovery of the first five representatives of a `new' population of galaxies, 
the ultra-compact dwarfs (UCDs), in the Fornax cluster. These have stellar
appearance in the DuPont plates used by \citet{ferguson89} in the Fornax 
Cluster Catalog (FCC), but nevertheless lie at the cluster redshift (see also 
Hilker et al. 1999). Deeper imaging and spectroscopic studies have now 
revealed populations of fainter UCDs in the Fornax (\citealt{mieske04a};
Drinkwater et al. 2005, in preparation) and Virgo \citep{drinkwater04,
jones05} clusters, as well as some likely UCDs in Abell 1689 
\citep{mieske04b}.

The nature of UCDs is still uncertain, but one intriguing possibility,
suggested by their apparent rarity in the field (Liske et al. 2005, in
preparation), is that they are the product of the cluster environment. 
Clusters are known to harbor unusual galaxy populations, rarely encountered 
elsewhere. One conspicuous example is provided by the class of `nucleated' 
dwarfs (dE,N) originally discovered in the Virgo cluster \citep{binggeli85}, 
which represent a large fraction of dwarfs in clusters but have far fewer 
counterparts in nearby groups (in the Local Group only NGC205 and the 
Sagittarius dwarf may be regarded as nucleated). Bekki et al. (2001, 2003) 
proposed that stripping of the low surface brightness envelopes around 
nucleated dwarfs may produce a compact remnant, whose properties would be 
comparable to UCDs (``galaxy threshing'').

A test of this hypothesis can be made by comparing the structural
properties of UCDs and dE,N nuclei. We present here the results of 
such a study carried out using images from the Hubble Space Telescope
(HST). We adopt Virgo and Fornax distance moduli of 30.92 and 31.39 
mag., respectively, from Cepheid distances \citep{freedman01} and 
extinctions derived from the maps of \citet{schlegel98}.

\section{Observations and Data Analysis}

The five original Fornax UCDs (\citet{drinkwater00},\citet{phillipps01}) 
and one Fornax nucleated dwarf (FCC303) were observed with the Space 
Telescope Imaging Spectrograph (STIS -- Woodgate et al.\ 1998) through 
the `open' (50CCD) filter in Cycle~8 (GO-8685). Exposure times were 
1680 seconds, broken down into at least 5 dither positions for cosmic 
ray removal and to better sample the point spread function.  We also 
use images of Virgo nucleated dwarfs taken as part of the ACS Virgo 
cluster survey (ACSVCS -- C\^ot\'e et al. 2004) and observed with the 
HST Advanced Camera for Surveys (ACS) through the $g$ and $z$ filters, 
with total exposure times of 750 and 1120 seconds, respectively. These 
images were retrieved as fully processed, drizzled and sky subtracted 
files from the HST archive\footnote{\tt http: //archive.stsci.edu} and 
analyzed as described below.

For STIS images of the UCDs, we ran the IRAF task `ellipse' \citep{jedr87}
to derive the surface brightness profile, which we calibrated on to the
AB system, with a correction to place the open filter data on the $V$
filter scale, assuming a solar spectral energy distribution, as specified
in the STIS online manual. The 50CCD bandpass is very broad and this makes
the transformation dependent on the underlying spectral energy distribution.
Most realistic spectra, however, do not have a significantly different 
correction from our assumed solar value.

The profiles are presented in Figure~1. Most UCDs have exponential
profiles with very similar scale lengths and central surface brightnesses;
only UCD3 is markedly different from all other objects, in that it is
much more extended and has a flatter central core than the other UCDs:
this object is fitted reasonably well by a \citet{sersic68} profile,
with (possibly) a small central excess. 

\begin{figure*}
\plottwo{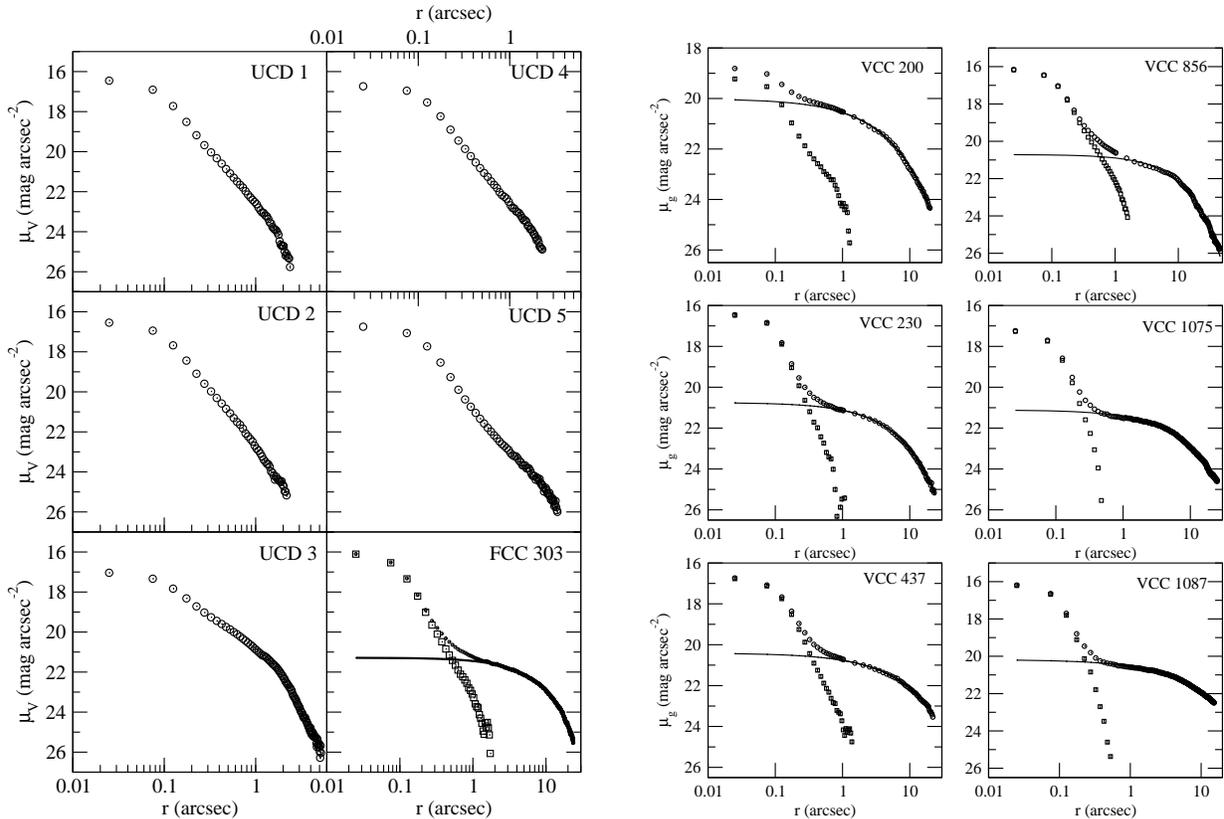}{f1b.eps}
\caption{Surface brightness profiles (circles) from STIS observations of
         the five original Fornax UCDs and FCC303 and a selection
         of dE,N from ACS imaging. The resolution of these profiles
         is $0.05''$. We show the Sersic model fitted to the envelopes 
         of the dE,N (solid line) and the nuclear profile after subtraction 
         of the model envelope (open squares). For clarity we only show one
         point in 5 of the dE,N envelopes, outside of the inner $1''$.}
\end{figure*}

We carried out the same procedure for the STIS image of FCC303 and for 
the ACS images of the dE,N.  We modelled the envelope of these galaxies 
using a \citet{sersic68} profile, regarded as the best model for such 
objects \citep{davies88,young94}.  We removed the model halo profile from
the data, leaving only the photometric residual of the nucleus.  Figure~1 
also shows the results of this procedure for FCC303 (observed with STIS) 
and a selection of the Virgo dE,N observed with ACS.  We will use these data 
to compare the properties of nuclei and UCDs.

As part of this process we also determine the luminosities and colors of 
the nuclei, using a $1''$ aperture from which we remove the flux from the 
underlying galaxy by integrating the Sersic profiles derived above. We use 
the ISHAPE software \citep{larsen99}, assuming a circular Plummer profile 
\citep{geha02} and using a point spread function generated by TinyTim 
\citep{krist97}, to determine half-light radii for the nuclei. We also 
derive the color for the host galaxy in the inner $1''$. These data are 
tabulated in Table 1. We put these data on the more commonly used Vega
system, rather than the AB system used in the figures (conversion of STIS open
filter data to the Vega system is not straightforward and requires 
assumptions as to the underlying spectral energy distribution; hence, 
we prefer to use AB values for these comparisons). Errors  for galaxy 
properties are derived from the Sersic profiles and errors for nuclear 
quantities are determined by assuming that the counts obey Poisson 
statistics and adding errors in quadrature. Errors for the size of the 
nucleus are difficult to determine, but \citet{larsen99} suggests that 
the half-light radii should be accurate to about 10\% for data of good 
quality.

\section{A comparison of nuclear and UCD structure}

Table 1 lists the properties of nuclei: absolute $g$ magnitude (on the
Vega system), $g-z$ color, half-light radius (from the Plummer
profile) and the $g-z$ color of the center of the host galaxy. We also
tabulate the $V$ magnitude of the UCDs, calculated over a large
aperture whose size is determined with reference to the profiles shown
in Figure 1, and their half-light radius for a Plummer profile (except
for UCD Fornax 3, which resembles a S\'ersic profile more
closely). The properties of the nuclei (other than their magnitude and
structure) are of secondary importance for our discussion, but these
dE,N nuclei resemble bright globular clusters \cite{harris96} while
the nuclei are somewhat bluer than the underlying galaxy, as found by
\cite{lotz04} for a sample observed with WFPC2.

\begin{deluxetable}{ccccc}
\tablecaption{Properties of nuclei and ultra-compact dwarfs}
\tabletypesize{\small}
\tablewidth{0.5\textwidth}
\tablehead{
\colhead{VCC \#} & \colhead{$M_g$ (nucleus)}   & \colhead{$g-z$ (nucleus)}
 & \colhead{$r_h$ (pc)} & \colhead{$g-z$ (galaxy)}
}
\startdata
 200 & $ -9.09 \pm 0.04$ & $ 1.23 \pm 0.06$ & 20.8 & $ 1.89 \pm 0.01$ \\
 230 & $-10.76 \pm 0.03$ & $ 1.65 \pm 0.05$ &  3.6 & $ 1.83 \pm 0.01$ \\
 437 & $-11.09 \pm 0.02$ & $ 1.62 \pm 0.04$ &  8.5 & $ 1.89 \pm 0.01$ \\
 856 & $-11.87 \pm 0.02$ & $ 1.86 \pm 0.03$ &  8.5 & $ 1.60 \pm 0.01$ \\
1075 & $ -9.62 \pm 0.04$ & $ 1.54 \pm 0.07$ &  4.0 & $ 1.80 \pm 0.01$ \\
1087 & $-10.60 \pm 0.03$ & $ 1.90 \pm 0.04$ &  2.3 & $ 1.94 \pm 0.01$ \\
1185 & $-10.19 \pm 0.04$ & $ 1.50 \pm 0.06$ &  4.4 & $ 1.91 \pm 0.01$ \\
1261 & $-11.40 \pm 0.03$ & $ 1.80 \pm 0.04$ &  4.6 & $ 1.80 \pm 0.01$ \\
1355 & $ -9.84 \pm 0.04$ & $ 1.71 \pm 0.06$ &  2.1 & $ 1.79 \pm 0.01$ \\
1407 & $-10.60 \pm 0.03$ & $ 1.59 \pm 0.04$ & 12.2 & $ 1.85 \pm 0.01$ \\
1431 & $-11.24 \pm 0.02$ & $ 1.55 \pm 0.03$ & 19.8 & $ 2.05 \pm 0.01$ \\
1489 & $ -8.32 \pm 0.08$ & $ 1.43 \pm 0.09$ &  4.4 & $ 1.65 \pm 0.01$ \\
1539 & $ -9.61 \pm 0.04$ & $ 1.58 \pm 0.05$ & 11.4 & $ 1.71 \pm 0.01$ \\
1826 & $-11.04 \pm 0.02$ & $ 1.34 \pm 0.03$ &  5.4 & $ 1.94 \pm 0.01$ \\
1886 & $ -8.61 \pm 0.08$ & $ 1.58 \pm 0.10$ &  3.3 & $ 1.58 \pm 0.01$ \\
1910 & $-11.07 \pm 0.02$ & $ 1.47 \pm 0.02$ &  4.6 & $ 2.01 \pm 0.01$ \\
2019 & $-11.06 \pm 0.03$ & $ 1.74 \pm 0.05$ &  2.3 & $ 1.81 \pm 0.01$ \\
2050 & $ -8.80 \pm 0.06$ & $ 1.17 \pm 0.09$ &  8.1 & $ 1.81 \pm 0.01$ \\
\hline \\
UCD  & $ M_V $             &                  &    &                  \\
\hline \\
Fornax 1    & $ -11.70 \pm 0.01$  &                  & 17.9 &                 \\
Fornax 2    & $ -11.79 \pm 0.01$  &                  & 20.3 &                 \\
Fornax 3    & $ -13.24 \pm 0.01$  &                  &      &                 \\
Fornax 4    & $ -11.90 \pm 0.01$  &                  & 20.6 &                 \\
Fornax 5    & $ -11.61 \pm 0.01$  &                  & 13.4 &                 \\

 \enddata

%% Text for table notes should follow after the \enddata but before
%% the \end{deluxetable}. Make sure there is at least one \tablenotemark
%% in the table for each \tablenotetext.

\end{deluxetable}

Figure 2 plots the surface brightness profiles of a selection of the
nuclei (the largest and smallest objects as well as those presented in
Figure 1) and the mean profile of the Fornax UCDs (excluding \#3).
These profiles were scaled to parsecs, based on published Cepheid
distances to the Fornax and Virgo clusters. The profiles are not
deconvolved, but the PSF is only relevant for the inner $0.2''$ while
we compare structures on scales of $1''$--$2''$. None of the nuclei,
with the possible exception of VCC856 (IC3328) and FCC303 (and VCC1431
which is not shown), is similar to the mean UCD profile.  Even for the
nuclei of VCC856, VCC1431 and FCC303's nuclei, however, there is a
discrepancy at large radii, where UCDs are more extended. In general,
nuclei appear to have smaller sizes, lower surface brightnesses, and,
to the extent that this is can be determined given the effects of the
point spread function, steeper profiles than UCDs: UCDs have
half-light radii of about 20 pc (cf. Drinkwater et al. 2003), while
nuclei have typical sizes of less than 10 pc. While the UCDs are
generally more luminous than nuclei, even nuclei of similar luminosity to
the UCDs tend to be physically smaller. UCD3 shows a more extended light
distribution but broadly resembles the other UCDs at smaller radii, 
suggesting that it may represent a transitional object between normal
dwarf ellipticals and UCDs.

\begin{figure}
\plotone{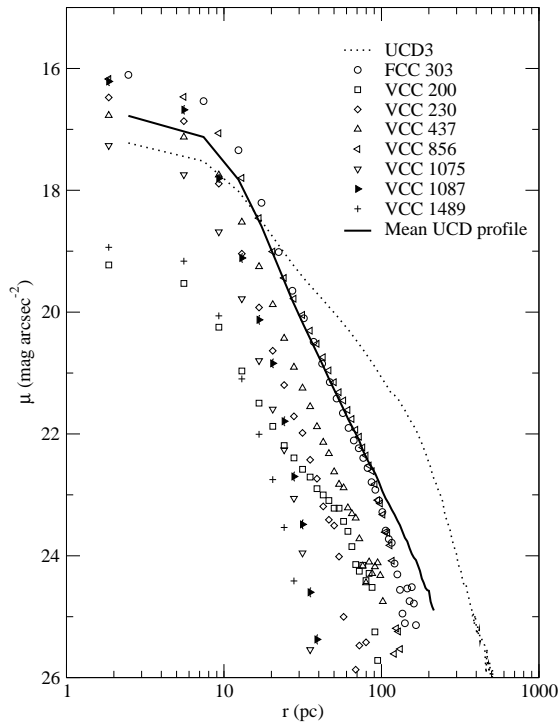}
\epsscale{0.85}
\caption{Comparison of surface brightness profiles for UCDs and the nuclei 
of dwarf ellipticals.}
\end{figure}

\citet{bekki01,bekki03} model the formation of UCDs by removal of the 
envelope from dwarf elliptical galaxies through tidal interactions with 
the gravitational potentical of the central galaxy.  These simulations 
show that the remnant nuclei properties are not significantly changed by the
threshing process (e.g. Figure 4 in Bekki et al. 2001).  If such is
the case, the Fornax UCDs studied here cannot be formed by the
extraction of nuclei from typical dE galaxies analyzed here.

The only exception to this appears to be VCC856. This dwarf galaxy is
known to possess a system of weak spiral arms and, possibly, a central
bar \citep{jerjen00}: VCC856 may be a dwarf example of the class of
anemic spirals \citep{vandenbergh76} encountered in clusters. The
possibility that UCDs represent remnant nuclei of dwarf spirals rather
than dwarf ellipticals has been mentioned by \citet{phillipps01} and
the similarity between the nucleus of VCC856 and UCDs offers some
support to this hypothesis (e.g., Moore et al. 1996).  VCC1431 and
FCC303, however, do not possess the obvious spiral structure observed
in VCC856.

One caveat is that the nuclei observed in the ACS Virgo Cluster Survey
may be biased somehow to smaller objects, although \citet{cote04}
state that the survey is designed to observe 100 galaxies fairly drawn
from the Virgo Cluster Catalog of \cite{sandage85}. The colors and
luminosities of the nuclei presented in Table 1 are consistent with
the larger sample of \citet{lotz04}. We have also determined sizes for
these latter nuclei and find that the size distribution for nuclei
observed with WFPC2 is consistent with the one presented in Table
1. This argues that the ACSVCS samples the population of dE nuclei
without obvious bias.

It is possible that the brighter Fornax UCDs are not proper 
representatives of the general UCD population; their similar structures 
and surface brightnesses suggest that we may be observing an extreme 
sample (the `tip of the iceberg' effect) while the fainter UCDs now 
discovered in the Fornax and Virgo clusters may be closer counterparts 
to the nuclei of typical surviving dE. A series of HST ACS HRC 
(high resolution camera) snapshots for these objects is now being 
acquired and will be discussed in a future paper.

The 5 UCDs in this study are the brightest known in Fornax, and the
real possibility exists that they have been selectively descended from
dE with particularly bright nuclei.  A more extensive sample of UCDs
and dEs imaged with HST can explore their connection more fully.  To
this end, we now have an ongoing HRC High Resolution Camera snapshot
survey of UCDs in both Fornax and Virgo; this should produce a sample
of 25 objects reaching 1-2 magnitudes fainter than the small STIS
sample.  The simulations carried out to date may not have sufficient
resolution to model the transition of a dE in a cluster environment
accurately, and perhaps there are additional physical processes which
come into play to cause the excess light at large radii in the UCDs.
Progress will be made via the interplay of the deeper and multicolor
datasets now in progress with HST in both the Fornax and Virgo
clusters.  These more extensive samples can drive improved modeling,
perhaps providing definitive tests for understanding the origin of
UCDs.

\acknowledgments

RDP is supported by a grant from the UK PPARC. MJD, EE and KB acknowledge
support from the Australian Research Council. MDG gratefully acknowledges 
support provided by NASA through grant number GO-8685 from the Space 
Telescope Science Institute, which is operated by AURA, Inc., under NASA 
contract NAS5-26555, and from NSF grant AST-04-07445.  Part of this work 
was performed under the auspices of the U.S. Department of Energy by 
University of California Lawrence Livermore National Laboratory under 
contract No.~W-7405-Eng-48. This work has made use of the archival
facilities at the Space Telescope Science Institute.

\clearpage


%% If you are not including electonic art with your submission, you may
%% mark up your captions using the \figcaption command. See the 
%% No more than seven \figcaption commands are allowed per page, 
%% so if you have more than seven captions, insert a \clearpage 

%% Tables should be submitted one per page, so put a \clearpage before

%% Two options are available to the author for producing tables:  the
%% deluxetable environment provided by the AASTeX package or the LaTeX
%% table environment.  Use of deluxetable is preferred.
%%

%% Three table samples follow, two marked up in the deluxetable environment,
%% one marked up as a LaTeX table.

%% In this first example, note that the \tabletypesize{}
%% command has been used to reduce the font size of the table.
%% Note also that the \label command needs to be placed 


\begin{thebibliography}{}

\bibitem[Arp(1965)]{arp65} Arp, H. C. 1965, \apj, 142, 402
\bibitem[Bekki et al.(2001)]{bekki01} Bekki, K., Couch, W. J. and 
    Drinkwater, M. J. 2001, \apj, 552, L105
\bibitem[Bekki et al.(2003)]{bekki03} Bekki, K., Couch, W. J., Drinkwater,
    M. J. and Shioya, Y. 2003, \mnras, 344, 399
\bibitem[Binggeli et al.(1985)]{binggeli85} Binggeli, B., Sandage, A. and
    Tammann, G. A. 1985, \aj, 90, 1681
\bibitem[Bothun et al.(1987)]{bothun87} Bothun, G., Impey, C. D.,
    Malin, D. and Mould, J. R. 1987, \aj, 94, 23
\bibitem[C\^ot\'e et al.(2004)]{cote04} C\^ot\'e, P. et al. 2004, \apjs, 153, 223
\bibitem[Cross et al.(2001)]{cross01} Cross, N. et al. 2001, \mnras, 324, 825
\bibitem[Dalcanton et al.(1997)]{dalcanton97} Dalcanton, J. J., Spergel, N.
    and Summers, F. J. 1997, \apj, 482, 659
\bibitem[Davies et al.(1988)]{davies88} Davies, J. I., Phillipps, S., 
    Cawson, M. G. M., Disney, M. and Kibblewhite, E. J. 1988, \mnras,
    232, 239
\bibitem[Drinkwater et al.(1999)]{drinkwater99} Drinkwater, M. J., 
    Phillipps, S., Gregg, M. D., Parker, Q. A., Smith, R. M., Davies, J. I.,
    Jones, J. B. and Sadler, E. M. 1999, \apj, 511, L97
\bibitem[Drinkwater et al.(2000)]{drinkwater00} Drinkwater, M. J., Jones,
    J. B., Gregg, M. D. and Phillipps, S. 2000, Pub. Ast. Soc. Australia,
    17, 227
\bibitem[Drinkwater et al.(2003)]{drinkwater03} Drinkwater, M. J., Gregg, 
    M. D., Hilker, M., Bekki, K., Couch, W. J., Ferguson, H. C., Jones, J. B.,
    and Phillipps, S. 2003, {\it Nature}, 423, 519
\bibitem[Drinkwater et al.(2004)]{drinkwater04} Drinkwater, M. J., Gregg,
    M. D., Couch, W. J., Ferguson, H. C., Hilker, M., Jones, J. B., Karick,
    A. and Phillipps, S. 2004, Pub. Ast. Soc. Australia, 21, 375
\bibitem[Disney(1976)]{disney76} Disney, M. 1976, {\it Nature}, 263, 573
\bibitem[Ferguson(1989)]{ferguson89} Ferguson, H. C. 1989, \aj, 98, 367
\bibitem[Freedman et al.(2001)]{freedman01} Freedman, W. C. et al. 2001, 
    \apj, 553, 47
\bibitem[Geha et al.(2002)]{geha02} Geha, M., Guhathakurta, P. and van
    der Marel, R. P. 2002, \aj, 124, 3073
\bibitem[Harris(1996)]{harris96} Harris, W. E. 1996, \aj, 112, 1487
\bibitem[Hilker et al.(1999)]{hilker99} Hilker, M., Infante, L.,
    Vieira, G., Kissler-Patig, M. and Richtler, T. 1999,\aaps, 134, 75
\bibitem[Impey et al.(1988)]{impey88} Impey, C., Bothun, G. and 
    Malin, D. 1988, \apj, 330, 634
\bibitem[Jedrzejewski(1987)]{jedr87} Jedrzejewski, R. I. 1987, \mnras,
    226, 747
\bibitem[Jerjen et al.(2000)]{jerjen00} Jerjen, H., Kalnajs, A. and 
    Binggeli, B. 2000, \aap, 358. 845
\bibitem[Jones et al.(2005)]{jones05} Jones, J. B. et al. 2005, \mnras, submitted
\bibitem[Krist and Hook(1997)]{krist97} Krist, J. E. and Hook, R. N.
    1997, in {\it The 1997 HST Calibration Workshop with a New
    Generation of Instruments}, p. 192
\bibitem[Kunth et al.(1988)]{kunth88} Kunth, D., Maurogordato, S. and
    Vigroux, L. 1988, \aap, 204, 10
\bibitem[Larsen(1999)]{larsen99} Larsen, S. S. 1999, \aaps, 139, 393
\bibitem[Lotz et al.(2004)]{lotz04} Lotz, J. M., Miller, B. C. and
    Ferguson, H. C. 2004, \apj, 613, 262
\bibitem[Mieske et al.(2004a)]{mieske04a} Mieske, S., Hilker, M. and 
    Infante, L. 2004a, \aap, 418, 445
\bibitem[Mieske et al.(2004b)]{mieske04b} Mieske, S. et al. 2004b, \aj,
    128, 1529
\bibitem[Mo et al.(1998)]{mo98} Mo, S., Mao, H. J. and White, S. D. M.
    1998, \mnras, 297, 71
\bibitem[Moore et al.(1996)]{moore96} Moore, B., Katz, N., Lake, G.,
     Dressler, A., Oemler, A. 1996, {\it Nature}, 379, 613
\bibitem[Phillipps et al.(2001)]{phillipps01} Phillipps, S., Drinkwater,
    M. J., Gregg, M. D. and Jones, J. B. 2001, \apj, 560, 201
\bibitem[Sandage et al.(1985)]{sandage85} Sandage, A., Binggeli, B. and
Tammann, G. A. 1985, \aj, 90, 1681
\bibitem[Schlegel et al.(1998)]{schlegel98} Schlegel, D. J., Finkbeiner,
    D. P. and Davis, M. 1998, \apj, 500, 525
\bibitem[S\'ersic(1968)]{sersic68} S\'ersic, J. 1968, Atlas de Galaxias
    Australes (Cordoba: Observatorio Astronomico)
\bibitem[van den Bergh(1976)]{vandenbergh76} van den Bergh, S. 1976,
    \apj, 206, 883
\bibitem[Woodgate et al.(1998)]{woodgate98} Woodgate, B. E. et al. 1998, \pasp, 110, 1183
\bibitem[Young and Currie(1994)]{young94} Young, C. K. and Currie, 
    M. J. 1994, \mnras, 268, L11

\end{thebibliography}
\end{document}